\documentclass[10pt,a4paper,english,preview]{revtex4-2}
\usepackage[T1]{fontenc}
\usepackage[latin9]{inputenc}
\usepackage{xcolor}
\usepackage{pdfcolmk}
\usepackage{verbatim}
\usepackage{prettyref}
\usepackage{amsmath}
\usepackage{amssymb}
\usepackage{cancel}
\usepackage{esint}
\PassOptionsToPackage{normalem}{ulem}
\usepackage{ulem}

\makeatletter

\pdfpageheight\paperheight
\pdfpagewidth\paperwidth

\providecolor{lyxadded}{rgb}{0,0,1}
\providecolor{lyxdeleted}{rgb}{1,0,0}

\DeclareRobustCommand{\lyxsout}[1]{\ifx\\#1\else\sout{#1}\fi}

\makeatother

\usepackage{babel}
\begin{document}
\title{On the derivation of guiding center dynamics without coordinate dependence}
\author{Zhi YU }
\affiliation{University of Science and Technology of China, Hefei, People's Republic
of China, 230027}
\affiliation{Institute of Plasma Physics, Hefei Institutes of Physical Science,
Chinese Academy of Sciences, Hefei, People's Republic of China, 230031}
\begin{abstract}
The fundament of the classical guiding center theory is gyro-phase
averaging, which cannot be well defined over a non-trivial magnetic
field topology. The local gyro-phase coordinate frame hides the geometric
nature of gyro-symmetry. A coordinate-free geometric representation
should be a more appropriate alternative for a deeper understanding
of the guiding center dynamics. In this paper, the motion of a charged
particle is described by a Lagrangian one-form on a seven-dimensional
phase space. The Lagrangian one-form is geometrically decomposed by
constructing a coordinate-free gyro-averaging method. As a result,
we obtain the coordinate-free expression of the non-relativistic guiding-center
dynamics in the time-dependent slow-varying electromagnetic field.
\end{abstract}
\maketitle

\section{Introduction}

The guiding center dynamics has been a subject of interest to researchers
for decades \citep{caryHamiltonianTheoryGuidingcenter2009}. The purpose
of guiding center theory is to decompose particle motion into vertical
gyro-motion and horizontal drift motion. The classical decomposition
method is based on the averaging of the gyro-phase. The gyro-phase
is defined on a predefined local orthogonal coordinate frame, which
may not exist globally over a non-trivial magnetic field. This issue
was raised and discussed by Sugiyama \citep{sugiyamaResponseCommentGuiding2009,sugiyamaGuidingCenterPlasma2008}
and Krommes \citep{krommesCommentGuidingCenter2009} in 2009. Soon
after, Burdy and Qin \citep{burbyGyrosymmetryGlobalConsiderations2012}
recognized that the magnetic field inhomogeneity obstructs the existence
of the global gyro-phase. The global gyro-phase is not a necessary
condition for gyro-symmetry. The gyro-symmetry depends only on the
homogeneity of the electromagnetic field within the range of gyro-motion.
The local gyro-phase coordinate frame hides the geometric nature of
gyro-symmetry and prevents us from identifying this issue. A coordinate-free
geometric representation should be a more appropriate alternative
for a deeper understanding of the guiding center dynamics. A series
of recent works by Burby on slow manifolds of near-periodic Hamiltonian
systems has shown the importance of geometric tools for understanding
gyro-symmetry \citep{burbyGuidingCenterDynamics2020,burbyGeneralFormulasAdiabatic2020,burbyNormalStabilitySlow2021}.

The base of guiding-center theory is the symmetry of trajectories
of the charged particles in electromagnetic fields, called \emph{gyro-symmetry}.
Kruskal \citep{kruskalAsymptoticTheoryHamiltonian1962} pointed out
that the set of charged particles with a common guiding center constitutes
a topological ring in phase space, called \emph{Kruskal's ring} \citep{qinShortIntroductionGeneral2005}.
In a slowly varying electromagnetic field, the charged particles on
the same Kruskal's ring have similar phase space trajectories. As
long as the particles do not resonate with the field, Kruskal's ring
will not break, but will only be slightly deformed. In other words,
the guiding center is the center of Kruskal's ring. The decomposition
of gyro-motion is to decompose the Kruskal's ring from the phase space
of particle trajectories. 

This paper aims to construct coordinate-free guiding-center dynamics,
or rather, Kruskal's ring dynamics. The description of Kruskal's ring
relies on two vector fields, the \emph{roto-rate} vector and the \emph{gyro-radius}.
The\emph{ }\textit{\emph{roto-rate}} vector is the generator of gyro-symmetry,
named by Kruskal \citep{kruskalAsymptoticTheoryHamiltonian1962}.
Omohundro\citep{omohundroGeometricPerturbationTheory1985} showed
the coordinate-free expression of the roto-rate vector. The gyro-radius
is the vector from the guiding center to the ringmates, representing
the mapping between Kruskal's ring and the guiding center. In an inhomogeneous
electromagnetic field, gyro-symmetry is not absolute. The gyro-decomposition
is an asymptotic approximation to the exact particle motion. The form
of gyro-radius is not unique -- different forms of gyro-radius correspond
to different decompositions \citep{parraEquivalenceTwoIndependent2014}.
We will show that a proper definition of the rote rate vector and
the gyro-radius may yield a concise expression of the guiding center
dynamics.

In this paper, we use the Lagrangian one-form (or called Poincar\'e-Cartan
one-form) to represent the motion of charge particle in a time-dependent
slow-varying electromagnetic field. The extended phase space of particle
is seven-dimensional contact manifold. The Lagrangian formalism provides
simple and explicit expressions for the variational principle and
Noether's theorem \citep{arnoldAppendixContactStructures1989}. From
the geometric point of view, the Lagrangian one-form is the dual counterpart
of the trajectory. The decomposition of the trajectory can be achieved
by decomposing the Lagrangian one-from. We will show that the non-existence
of global gyro-phase is a natural conclusion of the Lagrangian one-form
decomposition. As the result go geometric descomposition, we obtain
the coordinate-free expression of the non-relativistic guiding-center
dynamics in the time-dependent slow-varying electromagnetic field.

The derivation in this paper uses knowledge of elementary differential
geometry and Lie groups. For interested readers, Marsden and Ratiu's
book \citep{marsdenIntroductionMechanicsSymmetry1999} would be a
good reference. 

This paper is organized as follows. In \prettyref{sec:Geometric-expression},
we recall the general geometric setting of the Lagrangian formalism
for a time-dependent system and discuss the relation between Poincar\'e-Cartan
integral invariant and Noether's theorem. In \prettyref{sec:Kruskal's-ring},
we describe the decomposition of Kruskal's ring and the general expression
of the guiding center dynamics. The \prettyref{sec:Charged-particle-motion}
shows the coordinate-free expression of non-relativistic guiding center
dynamics in the time-dependent electromagnetic field. And, the \prettyref{sec:Summary}
is the summary and discussion.

\section{Geometric s etting \label{sec:Geometric-expression}}

Considering a time-dependent Hamiltonian system with the extended
phase space $P$, the action integral is an line integral along phase
space trajectory $\lambda$,
\begin{equation}
\mathcal{A}\left[\lambda\right]=\int_{\lambda}\eta\;,
\end{equation}
where one-form $\eta=pdq-Hdt$ is called the \emph{Lagrangian one-form}.
From Hamiltonian principle, the variation of trajectory $\lambda$
gives the \emph{Hamiltonian equations}

\begin{equation}
\iota_{\tau}d\eta=0\,,\label{eq:Hamiltonian equations}
\end{equation}
where $\iota_{\tau}$ means interior product with vector field $\tau$,
and $d\eta$ is the exterior derivative of one-form $\eta$.\textcolor{red}{{}
}The Hamiltonian flow is an one-parameter group of $t$ whose infinitesimal
generator is $\tau$, $\Psi_{t}^{H}=\exp\left(t\tau\right).$ If we
add a closed one-form to the Lagrangian one-form $\eta^{\prime}=\eta+\alpha$
and $d\alpha=0$, the result of Hamiltonian equations Eq.\prettyref{eq:Hamiltonian equations}
does not change. The extended phase space $P$ is an $2n+1$ dimensional
manifold\textcolor{red}{{} }endowed with a\emph{ }one-form $\eta$ that
satisfies the nonintegrable condition $\eta\wedge\left(d\eta\right)^{n}\neq0$.
This structure $\left(P,\eta\right)$ is a \emph{contact structure,
}and $\eta$ is also called \emph{contact form}. \textcolor{red}{}%
\textcolor{red}{{} }The Lagrangian one-form (contact one-form) $\eta$
plays an important role in time-dependent mechanics, which provides
simple and explicit expressions for the variational principles and
Noether's theorem \citep{arnoldAppendixContactStructures1989,marsdenIntroductionMechanicsSymmetry1999}.

\subsection{Poincar\'e-Cartan integral invariant}

Consider a curve $\mathcal{O}$ encircles a tube of phase trajectories
in extended phase space $P$, the action integral on $\mathcal{O}$
is denoted as $\mu_{P}\left[\mathcal{O}\right]\equiv\oint_{\mathcal{O}}\eta$.
Let $\mathcal{O}$ move along the same tube of phase trajectories,
the action integral on the image of Hamiltonian flow $\Psi_{t}^{H}$
looks like 
\begin{align}
\mu_{P}\left[\Psi_{t}^{H}\circ\mathcal{O}\right] & =\oint_{\Psi_{t}^{H}\circ\mathcal{O}}\eta=\oint_{\mathcal{O}}\Psi_{t}^{H*}\eta\nonumber \\
 & =\oint_{\mathcal{O}}\left(\eta+t\mathcal{L}_{\tau}\eta+\cdots\right)\nonumber \\
 & =\mu_{P}\left[\mathcal{O}\right]+t\oint_{\mathcal{O}}\left(\cancel{i_{\tau}d\eta}+di_{\tau}\eta\right)+\cdots\,.\label{eq:Poincare-Carten Equation}
\end{align}
Because $\iota_{\tau}d\eta=0$ and the circle integral of closed form
$di_{\tau}\eta$ is zero, the second and higher order terms of Eq.\prettyref{eq:Poincare-Carten Equation}
will vanish. Then, we can say that the action integral on a closed
phase space curve, $\mu_{P}\left[\mathcal{O}\right]$, is a constant
of motion. The one-form $\eta$ is also called the of \emph{Poincar\'e's
relative} \emph{integral invariant} or \emph{Poincar\'e-Cartan} one-form\citep{arnold44IntegralInvariang1989}.
The action integral $\mu_{P}\left[\mathcal{O}\right]$ is indepedent
with the shape of $\mathcal{O}$, which captures the topological property
of the bundle of trajectories. The Hamiltonian flow preserves the
action integral over arbitrarily closed loop in phase space. However,
without additional constraints, the Hamiltonian flow would not preserve
the compactness of the loop $\mathcal{O}$.

\subsection{Noether's theorem}

Consider an one-parameter Lie group $\Phi_{\theta s}=\exp\left(s\partial_{\theta}\right),s\in\mathbb{R}$,
which is generated by a vector field $\partial_{\theta}$. The action
integral of the infinitesimally transformed trajectory is 

\begin{align}
\mathcal{A}\left[\Phi_{\theta s}\circ\lambda\right] & =\int_{\Phi_{\theta s}\circ\lambda}\eta=\int_{\lambda}\Phi_{\theta s}^{*}\left(\eta\right)=\int_{\lambda}\exp\left(s\mathcal{L}_{\partial_{\theta}}\right)\eta=\mathcal{A}\left[\lambda\right]+s\int_{\lambda}\mathcal{L}_{\partial_{\theta}}\eta+O\left(s^{2}\right)\,,
\end{align}
where $\mathcal{L}_{\partial_{\theta}}$ is the Lie derivative along
the vector $\partial_{\theta}$. If $\mathcal{L}_{\partial_{\theta}}\eta=0$,
the higher order term of $s$ will vanish and action integral $\mathcal{A}\left[\lambda\right]$
is preserved under transformation of the group $\Phi_{\theta}$. We
shall say $\Phi_{\theta}$ is a \emph{Noether symmetry} on the Hamiltonian
system $\left(P,\eta\right)$. Using the Cartan's Magic Formula
\begin{align}
0=\mathcal{L}_{\partial_{\theta}}\eta & =\iota_{\partial_{\theta}}d\eta+d\iota_{\partial_{\theta}}\eta\;,\label{eq:Cartan's Magic Formula}
\end{align}
we get
\begin{equation}
\iota_{\partial_{\theta}}d\eta=-d\iota_{\partial_{\theta}}\eta=-d\mu\,,\label{eq:Noether theorem}
\end{equation}
where
\begin{equation}
\mu\equiv\iota_{\partial_{\theta}}\eta\,,
\end{equation}
is the \emph{moment map} induced by \emph{$\partial_{\theta}$} \citep{marsdenIntroductionMechanicsSymmetry1999}.
Putting $\iota_{\tau}$ on Eq. \prettyref{eq:Cartan's Magic Formula},
yields 

\begin{equation}
0=\iota_{\tau}\iota_{\partial_{\theta}}d\eta=-\iota_{\tau}d\iota_{\partial_{\theta}}\eta=-\iota_{\tau}d\mu
\end{equation}
It is easy to verify that $\mu$ is also an invariant under the action
of $\Phi_{\theta}$
\begin{equation}
\mathcal{L}_{\partial_{\theta}}\mu=0,\qquad\Phi_{\theta}^{*}\left(\mu\right)=\mu\,.
\end{equation}
Putting $\mathcal{L}_{\tau}$ on Eq. \prettyref{eq:Cartan's Magic Formula},
yields
\begin{equation}
\mathcal{L}_{\tau}\mathcal{L}_{\partial_{\theta}}\eta=\mathcal{L}_{\tau}\left(\iota_{\partial_{\theta}}d\eta+d\iota_{\partial_{\theta}}\eta\right)=i_{\left[\tau,\partial_{\theta}\right]}d\eta-\cancel{i_{\partial_{\theta}}\mathcal{L}_{\tau}d\eta}+d\mathcal{L}_{\tau}\left(i_{\partial_{\theta}}\eta\right)=0\,,
\end{equation}
where $\left[\cdot,\cdot\right]$ denotes the commutor of vector fields.
If $d\eta$ is not degenerate, the symmetry vector $\partial_{\theta}$
should commutes with Hamiltonian vector $\left[\tau,\partial_{\theta}\right]=0$.

Noether's theorem requires the Lie derivative of Poincar\'e-Cartan
one-form $\eta$ along the symmetry vector $\partial_{\theta}$ vanish,
which is only a local constraint on the one-form $\eta$. The Noether's
theorem Eq.\prettyref{eq:Noether theorem} can not tell us the global
topology of the symmetry. For the same Lie algebra $\partial_{\theta}$
, the orbit $\mathcal{O}$ of Lie group $\Phi_{\theta}$ may be isomorphic
to $S^{1}$ or $\mathbb{R}$. The global topology of the Lie group
$\Phi_{\theta}$ depends on the nature of the extended phase space
$\left(P,\eta\right)$. If the symmetry group $\Phi_{\theta s}$ is
a compact one-parameter group, it is isomorphic to the group $S^{1}\cong U\left(1\right)$,
whose orbit $\mathcal{O}$ is a closed curve\citep{hallLieGroupsLie2015}.
The action integral on the closed orbit $\mathcal{O}$ is

\begin{equation}
\mu_{P}\left[\mathcal{O}\right]=\oint_{\mathcal{O}}\eta=\int_{0}^{1}\Phi_{\theta s}^{*}\left(\iota_{\partial_{\theta}}\eta\right)ds=\mu\;.\label{eq:global vs local invariant}
\end{equation}
which is consistent with the moment map $\mu$. 

Noether's theorem determines the invariance of the moment map $\mu$.
In many cases, there is no guarantee that the exact symmetry always
exists everywhere in the phase space, and $\mu$ will not be a globally
valid exact invariant. If the inhomogeneity of the phase space is
a small quantity in the range of the closed loop $\mathcal{O}$, then
the variation of $\mu$ along the particle trajectory is bounded,
and $\mu$ is called \emph{adiabatic invariant} \citep{arnoldTheoryPerturbationsConditionally1989}.
Liouville's theorem determines the invariance of action integral $\mu_{P}$
on the closed loop $\mathcal{O}$, which is an absolute invariant.
Kruskal pointed out that points in the phase space of a near-periodic
dynamical system form closed loops that drift along phase space trajectories,
preserving their topology with only slight deformations \citep{kruskalAsymptoticTheoryHamiltonian1962}.
The phase space inhomogeneity leads to the deformation of $\mathcal{O}$,
and the deviation between $\mu$ and $\mu_{P}$. Since the deviation
is bounded, we use $\mu$ as the asymptotic approximation to $\mu_{P}$,
which can preserve the invariance to arbitrary orders\citep{burbyGeneralFormulasAdiabatic2020,burbyNormalStabilitySlow2021}.
We call these loops \emph{Kruskal's ring} or \emph{invariant tori}
in Arnold's book \citep{arnoldTheoryPerturbationsConditionally1989}.
The existence of Kruskal's ring implies that the divergent Hamiltonian
flow is constrained by a local compact group $\Phi_{\theta}$. Our
aim is to decompose Kruskal's ring $\mathcal{O}$ from the phase space
$P$ to obtain a quotient manifold $P/\mathcal{O}$.

\section{Kruskal's ring and guiding center\label{sec:Kruskal's-ring}}

\subsection{Kruskal's ring}

Consider the phase space $\left(P,\eta\right)$ and a local compact
Lie group $\Phi_{\theta}$ called \emph{gyro-transformation} or \emph{gyro-symmetry}.
If the Lagrangian one-form $\eta$ is invariant to the action of $\Phi_{\theta}$
throughout the phase space $P$, we say the phase space $P$ is \emph{uniform}
to the gyro-transformation $\Phi_{\theta}$. The orbit of $\Phi_{\theta}$
is called \emph{Kruskal's ring} $\mathcal{O}$, and the points on
the same orbit are called \emph{ringmates}, $\mathcal{O}_{z}=\left\{ \Phi_{\theta}\left(z\right)\right\} $.
The gyro-symmetry $\Phi_{\theta}$ induce a moment map $\mu:P\rightarrow\mathbb{R}$,
which projects Kruskal's ring to a constant of motion along the particle
trajectory. If the gyro-transformation has a fixed point $\Phi_{\theta}\left(Z\right)=Z$,
we called it the \emph{guiding center} and use it as the representative
of Kruskal's ring $\mathcal{O}$. 

The trajectories of Kruskal's ring constitute the quotient manifold
$P/\mathcal{O}$. The guiding center trajectory is isomorphic to the
Kruskal's ring trajectory $\bar{P}\simeq P/\mathcal{O}$. Then, a
natural projection arises 
\begin{equation}
\Pi_{-\rho}:P\rightarrow P/\mathcal{O}\times\mathcal{O}\rightarrow\bar{P}\times\mathcal{\bar{O}}\,,
\end{equation}
where the minus sign indicates the transformation is in the opposite
direction of the gyro-radius, and the superscript 'bar' indicates
that it is defined at the guiding center. Let inverse projection $\Pi_{\rho}$
be a one-parameter transformation generated by\emph{ gyro-radius $\Pi_{r\rho}=\exp\left(r\partial_{\bar{\rho}}\right)$,}
the pullback from ringmate to guiding center is a formal power series
of the Lie derivative $\mathcal{L}_{\partial_{\bar{\rho}}}$ 
\begin{equation}
\Pi_{\bar{\rho}r}^{*}\alpha=\sum_{n\ge0}\frac{r^{n}}{n!}\mathcal{L}_{\partial_{\bar{\rho}}}^{n}\left.\alpha\right|_{Z}\,,\label{eq:eta formal power series}
\end{equation}
where $\left.\alpha\right|_{Z}$ is the quantity defined at the guiding
center. 

Pushing forward $\partial_{\theta}$ to the guiding center, yields
the \emph{rote-rate vector }
\begin{equation}
\partial_{\bar{\theta}}\equiv\Pi_{-\rho*}\partial_{\theta}=\Pi_{\bar{\rho}}^{*}\partial_{\theta}\,,
\end{equation}
which is named by Kruskal \citep{kruskalAsymptoticTheoryHamiltonian1962}.
The orbit of rote-rate vector $\partial_{\bar{\theta}}$ is a circle
in the velocity space 
\begin{equation}
\Phi_{\bar{\theta}}=\exp\left(2\pi s\partial_{\bar{\theta}}\right)\leftrightarrow\bar{\mathcal{O}}\subset T_{Z}M\,,
\end{equation}
called \emph{limiting ring}. Therefore, we say the guiding center
is a particle with \textquotedbl spin\textquotedbl , whose magnetic
moment $\mu$ is equal to the action integral over the limiting ring
$\bar{\mathcal{O}}$. 

\subsection{Decomposition}

To decompose the motion of Kruskal's ring, we split the Hamiltonian
vector $\tau$ into two parts, the horizontal part $\tau_{\parallel}$
and the vertical part $\partial_{\theta}$,
\begin{align}
\tau & =\tau_{\parallel}+\partial_{\theta}\;.\label{eq:divide tau}
\end{align}
Substituting $\tau$ to Eq.\prettyref{eq:Hamiltonian equations},
yields the equation of horizontal motion
\begin{align}
\iota_{\tau_{\parallel}}d\eta & =-\iota_{\partial_{\theta}}d\eta=d\mu\;.\label{eq:horizontal Hamiltonian equation}
\end{align}
Pushing forward $\tau_{\parallel}$ to the guiding center, we get
the Hamiltonian vector field of guiding center
\begin{equation}
\bar{\tau}_{\parallel}=\Pi_{-\rho*}\tau_{\parallel}=\Pi_{\bar{\rho}}^{*}\tau_{\parallel}\,.
\end{equation}
\textit{\emph{It is straightforward to verify that the horizontal
motion is commute with vertical motions }}%
\begin{equation}
\left[\partial_{\theta},\tau\right]=\left[\partial_{\theta},\tau_{\parallel}\right]=0\;,
\end{equation}
and $\mu$ is a constant of motion in both directions
\begin{equation}
\iota_{\tau_{\parallel}}d\mu=\iota_{\partial_{\theta}}d\mu=0\,.
\end{equation}
 \textcolor{red}{}%
We also split the Lagrangian one-form into two parts

\begin{equation}
\eta=\eta_{\parallel}+\eta_{\perp}\,,
\end{equation}
\textit{\emph{ where the horizontal}} part is orthogonal to $\partial_{\theta}$
\begin{equation}
\iota_{\partial_{\theta}}\eta_{\parallel}=0\,,
\end{equation}
and \textit{\emph{the vertical}} part\textit{\emph{ gives the moment
map $\mu$}}
\begin{equation}
\iota_{\partial_{\theta}}\eta_{\perp}=\mu\,.
\end{equation}
From Eq. \eqref{eq:horizontal Hamiltonian equation}, the horizontal
vector field is given by
\begin{align}
*\tau_{\parallel}^{\flat} & =\frac{1}{2!\text{vol}_{\theta}}d\mu\wedge\eta_{\perp}\wedge d\eta\wedge d\eta\,,\label{eq:tau_horizontal}
\end{align}
where 
\begin{equation}
*\text{vol}_{\theta}\equiv\frac{1}{3!}\eta_{\perp}\wedge d\eta\wedge d\eta\wedge d\eta\,,
\end{equation}
is the phase space volume form introduced by $\eta_{\perp}$. Here,
the Hodge operator $*$ maps the $p$-form to $\left(n-p\right)$-form,
and the superscript $\flat$ means lowering of indices\citep{feckoDifferentialGeometryLie2006}.
Substituting Eq.\prettyref{eq:tau_horizontal} into Eq.\prettyref{eq:horizontal Hamiltonian equation},
one can verify that 

\begin{equation}
\iota_{\tau_{\parallel}}d\eta=d\mu-\cancel{\frac{\text{vol}_{\mu}}{\text{vol}_{\theta}}\eta_{\perp}}\,,
\end{equation}
 where
\begin{align}
\text{vol}_{\mu} & =*\left(\frac{1}{3!}d\mu\wedge d\eta\wedge d\eta\wedge d\eta\right)=\iota_{\tau}d\mu=0\,,
\end{align}
is the degenerate phase space volume form introduced by $d\mu$. 

\subsection{Perturbation}

The guiding center has practical significance only if the trajectories
of the ringmates are similar. In other words, the deformation of Kruskal's
ring should be limited. In the uniform phase space, Kruskal's ring
$\mathcal{O}$ is a circle. The deformation of Kruskal's ring came
from the inhomogeneity of phase space. Let a small quantity $\varepsilon\sim\mathcal{L}_{\partial_{\bar{\rho}}}$
denote the phase space inhomogeneity with respect to the gyro-radius
$\partial_{\bar{\rho}}$. In a slowly varying system, $\varepsilon\ll$1,
the deformation is a near-identity transformation generated by a perturbation
vector field $G$,
\begin{equation}
\Psi_{\varepsilon G}:\mathcal{O}\rightarrow\mathcal{O}_{\varepsilon},\quad,\Psi_{\varepsilon G}=\exp\left(\varepsilon G\right)\,.
\end{equation}
Further, we also have
\begin{equation}
\mathcal{O}_{\varepsilon}=\Psi_{\varepsilon G}\mathcal{O}=\Psi_{\varepsilon G}\circ\Pi_{\bar{\rho}}\circ\bar{\mathcal{O}}=\Pi_{\rho_{\varepsilon}}\circ\bar{\mathcal{O}}\;,
\end{equation}
where $\Pi_{\rho_{\varepsilon}}\equiv\Psi_{\varepsilon G}\circ\Pi_{\bar{\rho}}$
is the perturbed guiding-center projection. The perturbed gyro-transformation
$\Phi_{\varepsilon\theta}$ is given by 
\begin{equation}
\Phi_{\varepsilon\theta}\equiv\Pi_{\varepsilon\rho}\circ\Phi_{\bar{\theta}}\circ\Pi_{-\varepsilon\rho}=\Psi_{\varepsilon G}\circ\Pi_{\bar{\rho}}\circ\Phi_{\bar{\theta}}\circ\Pi_{-\bar{\rho}}\circ\Psi_{-\varepsilon G}=\text{Ad}_{\varepsilon G}\circ\text{Ad}_{\bar{\rho}}\circ\Phi_{\bar{\theta}}\;.
\end{equation}
Then, one can verify that the perturbed gyro-transformation $\Phi_{\varepsilon\theta}$
preserve the Noether's theorem
\begin{align}
\mathcal{\mathcal{A}}\left[\Phi_{\varepsilon\theta s}\circ\lambda_{\varepsilon}\right] & =\int_{\Phi_{\varepsilon\theta s}\circ\lambda_{\varepsilon}}\eta_{\varepsilon}=\int_{\lambda_{\varepsilon}}\Phi_{\varepsilon\theta s}^{*}\left(\eta_{\varepsilon}\right)\nonumber \\
 & =\int_{\lambda_{\varepsilon}}\Pi_{-\rho_{\varepsilon}}^{*}\circ\Phi_{\bar{\theta}s}^{*}\circ\Pi_{\rho_{\varepsilon}}^{*}\left(\eta_{\varepsilon}\right)\nonumber \\
 & =\int_{\lambda_{\varepsilon}}\Pi_{-\rho_{\varepsilon}}^{*}\circ\exp\left(s\mathcal{L}_{\partial_{\bar{\theta}}}\right)\circ\Pi_{\rho_{\varepsilon}}^{*}\left(\eta_{\varepsilon}\right)\nonumber \\
 & =\mathcal{A}\left[\lambda_{\varepsilon}\right]+s\int_{\Pi_{-\rho_{\varepsilon}}\circ\lambda_{\varepsilon}}\mathcal{L}_{\partial_{\bar{\theta}}}\Pi_{\rho_{\varepsilon}}^{*}\left(\eta_{\varepsilon}\right)+\cdots\nonumber \\
 & =\mathcal{A}\left[\lambda_{\varepsilon}\right]+s\int_{\bar{\lambda}}\mathcal{L}_{\partial_{\bar{\theta}}}\bar{\eta}+\cdots\;,
\end{align}
where the second and higher order terms will vanish at $\mathcal{L}_{\partial_{\bar{\theta}}}\bar{\eta}=0$.
Pullback $\bar{\mu}$ to the perturbed trajectory, the perturbed moment
map $\mu_{\varepsilon}=\iota_{\partial_{\theta\varepsilon}}\eta_{\varepsilon}$
is an constant of motion along the perturbed trajectory

\begin{equation}
0=\Pi_{-\rho_{\varepsilon}}^{*}\left(\iota_{\bar{\tau}}d\bar{\mu}\right)=\iota_{\Pi_{-\rho_{\varepsilon}}^{*}\bar{\tau}}d\Pi_{-\rho_{\varepsilon}}^{*}\bar{\mu}=\iota_{\tau_{\varepsilon}}d\mu_{\varepsilon}\;,
\end{equation}
Following the Lie perturbation method developed by Littlejohn\citep{littlejohnHamiltonianPerturbationTheory1982,littlejohnGuidingCenterHamiltonian1979}
and Cary \citep{caryNoncanonicalHamiltonianMechanics1983,caryLieTransformPerturbation1981}.
The perturbated moment map $\mu_{\varepsilon}$ may be calculated
to arbitrary order. 

The stable Kruskal's ring implies that ringmates' trajectories have
symmetry $\Phi_{\theta}\leftrightarrow\mathcal{O}$. The deformed
Kruskal's ring correspond to the perturbated gyro-symmetry $\Phi_{\theta\varepsilon}\leftrightarrow\mathcal{O}_{\varepsilon}$.
The action integral along the Kruskal's ring $\mu_{P}\left[\mathcal{O}_{\varepsilon}\right]$
is an exact invariant. As long as the symmetry group $\Phi_{\theta}$
is compact, the exact invariant $\mu_{P}$ always exist. Therefore,
we say the perturbed moment map $\mu_{\varepsilon}$ is an asymptotical
approximation of $\mu_{P}$. 

\subsection{Gyro-averaging}

The inhomogeneities in phase space break the gyro-symmetry. The purpose
of gyro-averaging is to eliminate the perturbation and to obtain the
gyro-invariant unperturbed Lagrangian. The gyro-averaging is an integral
along the Kruskal's ring
\begin{equation}
\left\langle \alpha\right\rangle =\int_{0}^{1}\Phi_{\theta s}^{*}\left(\alpha\right)ds\,.
\end{equation}
Pulling it back to the limiting ring, yields 

\begin{align}
\left\langle \alpha\right\rangle  & =\int_{0}^{1}\Phi_{\bar{\theta}s}^{*}\circ\Pi_{\bar{\rho}\varepsilon}^{*}\left(\alpha\right)ds=\int_{0}^{1}\sum_{n\ge0}\frac{\varepsilon^{n}}{n!}\mathcal{L}_{\Phi_{\bar{\theta}s}^{*}\partial_{\bar{\rho}}}^{n}\Phi_{\bar{\theta}s}^{*}\left(\left.\alpha\right|_{Z}\right)ds\;.
\end{align}
If $\left.\alpha\right|_{Z}$ is gyro-independent, the gyro-average
$\left\langle \alpha\right\rangle $ only dependent on the rotation
of gyro-radius $\Phi_{\bar{\theta}s}^{*}\partial_{\bar{\rho}}$, which
is a formal power series of $\mathcal{L}_{\partial_{\bar{\theta}}}$
\begin{align}
\Phi_{\bar{\theta}s}^{*}\partial_{\bar{\rho}} & =\exp\left(2\pi s\mathcal{L}_{\partial_{\bar{\theta}}}\right)\left(\partial_{\bar{\rho}}\right)=\sum_{n\ge0}\frac{\left(2\pi s\right)^{n}}{n!}\mathcal{L}_{\partial_{\bar{\theta}}}^{n}\left(\partial_{\bar{\rho}}\right)\,.
\end{align}

Note that the form of gyro-radius $\partial_{\bar{\rho}}$ and rote-rate
vector $\partial_{\bar{\theta}}$ are not unique, as long as the mapping
from the limiting ring $\bar{\mathcal{O}}$ to the Kruskal's ring
$\mathcal{O}$ holds, 
\begin{equation}
\Pi_{\bar{\rho}}\bar{\mathcal{O}}\mapsto\mathcal{O}\leftrightarrow\partial_{\theta}=\Pi_{\bar{\rho}*}\partial_{\bar{\theta}}\,,
\end{equation}
To simplify the gyro-averaging, we let gyro-radius $\partial_{\bar{\rho}}$
be 'complex-like' under the action of rote-rate vector $\partial_{\bar{\theta}}$
\begin{equation}
\mathcal{L}_{\partial_{\bar{\theta}}}^{2}\partial_{\bar{\rho}}=-\partial_{\bar{\rho}}\,.\label{eq:complex-like condition}
\end{equation}
Then, the rotation of gyro-radius $\Phi_{\bar{\theta}s}^{*}\partial_{\bar{\rho}}$
is a simple trigonometric polynomial
\begin{align}
\Phi_{\bar{\theta}s}^{*}\partial_{\bar{\rho}} & =\partial_{\rho}+\left(2\pi s\right)\partial_{\check{\rho}}-\frac{1}{2}\left(2\pi s\right)^{2}\partial_{\rho}-\frac{1}{3!}\left(2\pi s\right)^{3}\partial_{\check{\rho}}+\cdots\nonumber \\
 & =\sum_{n\ge0}\frac{\left(-1\right)^{n}\left(2\pi s\right)^{2n}}{2n!}\partial_{\rho}+\sum_{n\ge0}\frac{\left(-1\right)^{n}\left(2\pi s\right)^{2n+1}}{\left(2n+1\right)!}\partial_{\check{\rho}}\nonumber \\
 & =\cos\left(2\pi s\right)\partial_{\rho}+\sin\left(2\pi s\right)\partial_{\check{\rho}}
\end{align}
where $\partial_{\check{\rho}}\equiv\mathcal{L}_{\partial_{\bar{\theta}}}\partial_{\bar{\rho}}$
is the dual vector orthogonal to $\partial_{\bar{\rho}}$. And, the
gyro-average $\left\langle \alpha\right\rangle $ is a polynomials
in $\mathcal{L}_{\partial_{\bar{\rho}}}$ and $\mathcal{L}_{\partial_{\check{\rho}}}$
\begin{align}
\left\langle \alpha\right\rangle  & =\sum_{n\ge0}\frac{1}{n!}\int_{0}^{1}ds\left(\cos\left(2\pi s\right)\mathcal{L}_{\partial_{\bar{\rho}}}+\sin\left(2\pi s\right)\mathcal{L}_{\partial_{\check{\rho}}}\right)^{n}\left.\alpha\right|_{Z}\,.\label{eq:gyro-average}
\end{align}

\section{Charged particle motion in a slow-varying electromagnetic field\label{sec:Charged-particle-motion}}

\subsection{Poincar\'e-Cartan-Einstein one-form}

The motion of charged particle was considered in the four-dimensional
Minkowski space $M=E^{1,3}$ with global Cartesian coordinates $x^{\alpha}\equiv\left(x^{0},x^{i}\right)=\left(ct,\mathbf{x}\right)$
and the metric tensors takes the form $g=\left(-1,1,1,1\right)$.
The phase space $P$ is the cotangent bundle $T^{*}M$ of $M$ with
constraint condition, 
\begin{equation}
P\equiv\left\{ \left.\left(x,p\right)\right|x\in M,p\in T_{x}^{*}M,g^{-1}\left(p,p\right)=-m^{2}c^{2}\right\} \,,
\end{equation}
where $p\equiv\left(p^{0},\mathbf{p}\right)$ is the four-momentum.
For a charged particle in an electromagnetic field the Lagrangian
one-form (also known as the Poincar\'e-Cartan-Einstein one-form \citep{qinPullbackTransformationsGyrokinetic2004}
) is given by
\begin{eqnarray}
\eta & = & A+p=\left(\mathbf{A}+\mathbf{p}\right)\cdot d\mathbf{x}-\left(\phi+p^{0}\right)dt\,,\label{eq:pce-one-form}
\end{eqnarray}
where $A^{\alpha}\equiv\left(\phi,\mathbf{A}\right)$ is the four-potential.
The natural unit system is adopted , let $m=c=e=1$, and only consider
the non-relativistic case, 
\begin{equation}
p^{0}=\frac{\mathbf{v}^{2}}{2},\;\mathbf{p}=\mathbf{v}\,.
\end{equation}
The Hamiltonian vector $\tau$ is solved from Hamiltonian equation
Eq.\prettyref{eq:Hamiltonian equations}, 
\begin{equation}
\begin{cases}
\frac{\tau_{\mathbf{x}}}{\tau_{t}}=\frac{d\mathbf{x}}{dt} & =\mathbf{v}\;,\\
\frac{\tau_{\mathbf{v}}}{\tau_{t}}=\frac{d\mathbf{v}}{dt} & =\mathbf{v}\times\mathbf{B}+\mathbf{E}\;,
\end{cases}
\end{equation}
where the electromagnetic fields are epxressed in terms of the potentials
as $\mathbf{B}=\nabla\times\mathbf{A}$ and $\mathbf{E}=-\nabla\phi-\partial_{t}\mathbf{A}$. 

Define two auxiliary vector fields, one is the unit vector along the
direction of the magnetic field
\begin{equation}
\mathbf{b}\equiv\frac{\mathbf{B}}{\left|B\right|}\;,
\end{equation}
and the other is the $\mathbf{E}\times\mathbf{B}$ drift velocity 

\begin{equation}
\mathbf{D}\equiv\frac{\mathbf{E}\times\mathbf{B}}{\mathbf{B}^{2}}\;.
\end{equation}
The velocity $\mathbf{v}$ is decomposed into horizontal part $\mathbf{v}_{\parallel}$
and vertical part $\mathbf{v}_{\perp}$

\begin{align}
\mathbf{v}_{\parallel} & \equiv\mathbf{v}\cdot\mathbf{b}\mathbf{b}+\mathbf{D}\;,\\
\mathbf{v}_{\perp} & \equiv\mathbf{v}-\mathbf{v}\cdot\mathbf{b}\mathbf{b}-\mathbf{D}\;,
\end{align}
If $\mathbf{D}\neq0$, $\mathbf{v}_{\perp}$ and $\mathbf{v}_{\parallel}$
are not orthogonal 
\begin{equation}
\mathbf{v}_{\parallel}^{2}+\mathbf{v}_{\perp}^{2}=\mathbf{v}^{2}-2\mathbf{v}_{\perp}\cdot\mathbf{D}\;.
\end{equation}
The decomposition of the Hamilitonian vector $\mathbf{\tau}=\tau_{\parallel}+\tau_{\perp}$
looks like 
\begin{align}
\tau_{\parallel} & =\mathbf{v}_{\parallel}\cdot\partial_{\mathbf{x}}+\mathbf{E}\cdot\mathbf{b}\mathbf{b}\cdot\partial_{\mathbf{v}}+\partial_{t}\;,\\
\tau_{\perp} & =\mathbf{v}_{\perp}\cdot\partial_{\mathbf{x}}+\mathbf{v}_{\perp}\times\mathbf{B}\cdot\partial_{\mathbf{v}}\;.\label{eq:tau_perp}
\end{align}

The gyro-radius in configuration space is a three-dimensional vector
field
\begin{equation}
\boldsymbol{\mathbf{\rho}}\equiv\frac{\mathbf{b}\times\mathbf{v}_{\perp}}{\left|B\right|}\;.\label{eq:gyro radius}
\end{equation}
Using $\boldsymbol{\rho}$, we decompose the the four-momentum $p$
into vertical part 
\begin{align}
p_{\perp} & \equiv-\iota_{\boldsymbol{\rho}\cdot\partial_{\mathbf{x}}}dA=\boldsymbol{\rho}\times\mathbf{B}\cdot d\mathbf{x}-\mathbf{E}\cdot\boldsymbol{\mathbf{\rho}}dt\nonumber \\
 & =\left(\mathbf{v}-\mathbf{v}\cdot\mathbf{b}\mathbf{b}-\mathbf{D}\right)\cdot d\mathbf{x}-\left(\mathbf{v}-\mathbf{D}\right)\cdot\mathbf{D}dt\nonumber \\
 & =\mathbf{v}_{\perp}\cdot\left(d\mathbf{x}-\mathbf{D}dt\right)\;,
\end{align}
and horizontal part
\begin{align}
p_{\parallel} & \equiv p-p_{\perp}\nonumber \\
 & =\left(\mathbf{v}\cdot\mathbf{b}\mathbf{b}+\mathbf{D}\right)\cdot d\mathbf{x}-\frac{\left(\mathbf{v}-\mathbf{D}\right)^{2}+\mathbf{D}^{2}}{2}dt\nonumber \\
 & =\mathbf{v_{\parallel}}\cdot d\mathbf{X}-\frac{\mathbf{v}_{\parallel}^{2}+\mathbf{v}_{\perp}^{2}}{2}dt\nonumber \\
 & =\mathbf{v_{\parallel}}\cdot d\mathbf{X}-\frac{\mathbf{v_{\parallel}}^{2}}{2}dt-\mu\left|B\right|dt\,.
\end{align}
Appending a closed form $-d\left(\boldsymbol{\rho}\cdot\mathbf{A}\right)$
to the Poincar\'e-Cartan-Einstein one-form Eq.\prettyref{eq:pce-one-form},
yields
\begin{align}
\eta & =A+p_{\parallel}-\mathcal{L}_{\boldsymbol{\rho}\cdot\partial_{\mathbf{x}}}A\,,\label{eq:pce_oneform_prepared}
\end{align}
which will simplify our subsequent derivation.

\subsection{Gyro-transformation}

First consider the case that the electromagnetic field is homogeneous,
$\varepsilon=0$. The gyro-transformation is a rotation of the gyro-radius
$\boldsymbol{\rho}$
\begin{equation}
\Phi_{\theta}\left(z\right)=\left(t,\mathbf{x}+\boldsymbol{\rho}\left(\cos\theta-1\right)+\boldsymbol{\rho}\times\mathbf{b}\sin\theta,\mathbf{v}+\left|B\right|\boldsymbol{\rho}\times\mathbf{b}\left(\cos\theta-1\right)-\left|B\right|\boldsymbol{\rho}\sin\theta\right)\,,
\end{equation}
whose generator is obtained from vertical Hamiltonian vector field
$\tau_{\perp}$ (Eq.\prettyref{eq:tau_perp})
\begin{equation}
\partial_{\theta}\equiv\frac{d}{ds}\left.\Phi_{\theta s}\right|_{s=0}=\boldsymbol{\rho}\times\mathbf{b}\cdot\partial_{\mathbf{x}}-\left|B\right|\boldsymbol{\rho}\cdot\partial_{\mathbf{v}}=\frac{\tau_{\perp}}{\left|B\right|}\;.\label{eq:D_theta}
\end{equation}
The action integral along the Kruskal's ring $\mathcal{O}$ is 
\begin{align}
\mu_{P}\left[\mathcal{O}\right] & =\frac{1}{2\pi}\oint_{\mathcal{O}}\left(A+p\right)\nonumber \\
 & =-\frac{1}{2\pi}\int_{D_{\mathcal{O}}}\mathbf{B}\cdot d\mathbf{S}+\frac{1}{2\pi}\int_{0}^{2\pi}\Phi_{\theta}^{*}\left(\iota_{\partial_{\theta}}p\right)d\theta\nonumber \\
 & =-\frac{\left|B\right|\rho^{2}}{2}+\frac{\mathbf{v}_{\perp}^{2}}{\left|B\right|}=\frac{\left|B\right|\rho^{2}}{2}=\frac{\mathbf{v}_{\perp}^{2}}{2\left|B\right|}\;,\label{eq:mu_action integral}
\end{align}
where $D_{\mathcal{O}}$ is the area enclosed by $\mathcal{O}$, and
the Stokes' theorem is applied. The Eq.\prettyref{eq:mu_action integral}
is the original definition of the \emph{magnetic moment} \citep{jacksonClassicalElectrodynamics1999}. 

The calculation of the moment map $\mu=\iota_{\partial_{\theta}}\eta$
is tricky because the potential $A\left(t,\mathbf{x}\right)$ of uniform
electromagnetic field are not constant. The traditional solution is
to expand $\eta$ around the guiding center $Z=\left(t,\mathbf{X},\mathbf{V}\right)$
with respect the gyro-radius $\boldsymbol{\rho}$\citep{kruskalAsymptoticTheoryHamiltonian1962}.
Expanding Eq. \prettyref{eq:pce_oneform_prepared} with respect $\boldsymbol{\rho}\cdot\partial_{\mathbf{x}}$,
yields

\begin{align}
\bar{\eta}=\left.\eta\right|_{X} & =p_{\parallel}+A+\mathcal{L}_{\boldsymbol{\rho}\cdot\partial_{\mathbf{x}}}A+\varepsilon\frac{1}{2}\mathcal{L}_{\boldsymbol{\rho}\cdot\partial_{\mathbf{x}}}^{2}A\nonumber \\
 & -\mathcal{L}_{\boldsymbol{\rho}\cdot\partial_{\mathbf{x}}}A-\varepsilon\mathcal{L}_{\boldsymbol{\rho}\cdot\partial_{\mathbf{x}}}^{2}A+O\left(\varepsilon^{2}\right)\nonumber \\
 & =A+p_{\parallel}-\frac{1}{2}\boldsymbol{\rho}\cdot d\mathbf{V}+O\left(\varepsilon\right)\,.\label{eq:eta_0}
\end{align}

Pushing $\partial_{\theta}$ forward to the guiding center, yields
the limiting rote-rate vector $\partial_{\bar{\theta}}$ as follows
\begin{equation}
\partial_{\bar{\theta}}=\Pi_{\left(-\boldsymbol{\rho}\cdot\partial_{\mathbf{x}}\right)*}\partial_{\theta}=-\left|B\right|\boldsymbol{\rho}\cdot\partial_{\mathbf{V}}+O\left(\varepsilon\right)\,.
\end{equation}
{} It is straightforward to verify that the Lagrangian one-form $\bar{\eta}$
(Eq. \prettyref{eq:eta_0}) is gyro-invariant in a uniform electromagnetic
field 
\begin{align}
\mathcal{L}_{\partial_{\bar{\theta}}}\bar{\eta} & =0\,,\\
\iota_{\partial_{\bar{\theta}}}d\bar{\eta} & =-d\iota_{\partial_{\bar{\theta}}}\bar{\eta}=-d\bar{\mu}\,.
\end{align}
and the magnetic moment $\mu$ is given by
\begin{equation}
\bar{\mu}=\iota_{\partial_{\bar{\theta}}}\bar{\eta}=\frac{\left|B\right|\rho^{2}}{2}\;,
\end{equation}
which is equal to the action integral along the Kruskal's ring $\mu_{P}\left[\mathcal{O}\right]$.

\subsection{Gyro-average}

For the inhomogeneous case $\varepsilon\neq0$, we need pull the Lagrangian
one-form $\eta$ back to the guiding center. Substituting Eq.\prettyref{eq:pce_oneform_prepared}
to Eq.\prettyref{eq:eta formal power series}, yields%
{} 

\begin{align}
\eta_{\varepsilon} & =\sum_{n\ge0}\frac{\varepsilon^{n}}{n!}\mathcal{L}_{\partial_{\rho}}^{n}\left(\frac{1}{\varepsilon}A+p_{\parallel}-\mathcal{L}_{\partial_{\rho}}A\right)=A+p_{\parallel}+\sum_{n>0}\frac{\varepsilon^{n}}{n!}\eta_{n}\,,\label{eq:expand_gamma}
\end{align}
where
\begin{align}
\eta_{n} & \equiv\mathcal{L}_{\partial_{\rho}}^{n}p_{\parallel}-\frac{n}{n+1}\mathcal{L}_{\partial_{\rho}}^{n+1}A\;.\label{eq:eta_n}
\end{align}
The head order does not depend on the gyro-radius $\partial_{\rho}$
\begin{equation}
\eta_{\parallel}=A+p_{\parallel}\,,
\end{equation}
which is the horizontal Lagrangian one-form. The other orders are
polynomials of Lie derivative $\mathcal{L}_{\partial_{\rho}}$
\begin{equation}
\eta_{\perp}=\sum_{n>0}\frac{\varepsilon^{n}}{n!}\eta_{n}\,,
\end{equation}
which form the vertical Lagrangian one-form. 

Since $p_{\parallel}$ and $A\left(t,\mathbf{X}\right)$ are gyro-independent,
the gyro-transformation of $\eta_{\varepsilon}$ only depend on the
rotation of gyro-radius $\partial_{\bar{\rho}}$

\begin{align}
\Phi_{\bar{\theta}s}^{*}\eta_{n} & =\mathcal{L}_{\Phi_{\bar{\theta}s}^{*}\partial_{\bar{\rho}}}^{n}p_{\parallel}-\frac{n}{n+1}\mathcal{L}_{\Phi_{\bar{\theta}s}^{*}\partial_{\bar{\rho}}}^{n+1}A\,.
\end{align}
Let the gyro-radius $\partial_{\bar{\rho}}$ satisfy the complex-like
condition Eq.\prettyref{eq:complex-like condition}, then the gyro-averaged
Lagrangian one-form $\left\langle \eta_{n}\right\rangle $ are polynomials
of the Lie derivative $\mathcal{L}_{\partial_{\bar{\rho}}}$ (see
Eq.\prettyref{eq:gyro-average}),
\begin{align}
\left\langle \eta_{n}\right\rangle  & =\int_{0}^{1}ds\left(\cos\left(2\pi s\right)\mathcal{L}_{\partial_{\bar{\rho}}}+\sin\left(2\pi s\right)\mathcal{L}_{\partial_{\check{\rho}}}\right)^{n}p_{\parallel}\nonumber \\
 & -\frac{n}{n+1}\int_{0}^{1}ds\left(\cos\left(2\pi s\right)\mathcal{L}_{\partial_{\bar{\rho}}}+\sin\left(2\pi s\right)\mathcal{L}_{\partial_{\check{\rho}}}\right)^{n+1}A\,.\label{eq:averaged_eta_n}
\end{align}

Let the\emph{ }rote-rate vector be 
\begin{equation}
\partial_{\bar{\theta}}\equiv-\left|B\right|\boldsymbol{\rho}\cdot\partial_{\mathbf{V}}\;.\label{eq:rote-rate vector}
\end{equation}
The gyro-radius is obtaineded from the complex-like condition Eq.\prettyref{eq:complex-like condition}
\begin{align}
\partial_{\bar{\rho}} & \equiv\boldsymbol{\rho}\cdot\partial_{\mathbf{X}}+\varepsilon\left(-\boldsymbol{\rho}\cdot\nabla\mathbf{b}\mathbf{\times}\mathbf{b}\mathbf{\times}\left(\mathbf{V}-\mathbf{D}\right)+\boldsymbol{\rho}\cdot\nabla\mathbf{D}\right)\cdot\partial_{\mathbf{V}}\,,\label{eq:rho}
\end{align}
whose second term come from the perturbation caused by the inhomogeneity.
There are different equivalent forms of $\partial_{\bar{\rho}}$ and
$\partial_{\bar{\theta}}$ that correspond to different decomposition
of gyro-motion \citep{parraEquivalenceTwoIndependent2014}. How to
find a suitable decomposition for the gyro-motion is a problem worthy
of in-depth discussion.

\subsection{The guiding-center dynamic\label{subsec:Gyro-radius-and-roto-rate} }

In order to obtain the dynamics of the guiding center, we first perform
the gyro-average on the Lagrangian one-form $\bar{\eta}=\left\langle \eta_{\varepsilon}\right\rangle $.
The vertical part is given by Eq.\prettyref{eq:averaged_eta_n}, 
\begin{align}
\bar{\eta}_{\perp} & =\left\langle \eta_{\perp}\right\rangle =-\frac{1}{4}\varepsilon\left(\mathcal{L}_{\partial_{\bar{\rho}}}^{2}A+\mathcal{L}_{\partial_{\check{\rho}}}^{2}A\right)+O\left(\varepsilon^{2}\right)\,.
\end{align}
{} Substituting Eq.\prettyref{eq:rho}, yields 

\begin{align}
\bar{\eta}_{\perp} & =-\frac{1}{2}\boldsymbol{\rho}\cdot d\mathbf{V}_{\perp}+\frac{\bar{\mu}}{2}\left(\mathbf{b}\cdot\nabla\times\mathbf{b}\right)\mathbf{b}\cdot d\mathbf{X}+\frac{\bar{\mu}}{2}\left(\mathbf{b}\cdot\nabla\times\mathbf{D}\right)dt\,,\label{eq:eta_perp}
\end{align}
 where 
\begin{align}
\bar{\mu} & \equiv\iota_{\partial_{\bar{\theta}}}\bar{\eta}=\iota_{\partial_{\bar{\theta}}}\bar{\eta}_{\perp}=\frac{\mathbf{V}_{\perp}^{2}}{2\left|B\right|}\;,
\end{align}
is the magnetic moment. 

Extracting the factor $\text{\ensuremath{\mu}}$ from $\bar{\eta}_{\perp}$,
yields a dimensionless one-form%
\begin{align}
\frac{\eta_{\perp}}{\mu}=\sigma & \equiv-d\mathbf{c}\cdot\mathbf{a}+\frac{1}{2}\left(\mathbf{b}\cdot\nabla\times\mathbf{b}\right)\mathbf{b}\cdot d\mathbf{X}-\frac{1}{2}\mathbf{b}\cdot\nabla\times\mathbf{D}dt\,,\label{eq:sigma}
\end{align}
where
\begin{align}
\mathbf{c} & \equiv\frac{\mathbf{V}_{\perp}}{\left|\mathbf{V}_{\perp}\right|}=\mathbf{a}\times\mathbf{b}\,,\\
\mathbf{a} & \equiv-\mathcal{L}_{\partial_{\bar{\theta}}}\mathbf{c}=\mathbf{b}\times\mathbf{c}=\frac{\mathbf{b}\times\mathbf{V}_{\perp}}{\left|\mathbf{V}_{\perp}\right|}\,,
\end{align}
are two unit vector fields perpendicular to the direction of magnetic
field $\mathbf{b}$. %
{} The dimensionless one-form $\sigma$ only depends on the spatial-temporal
inhomogeneity of the electromagnetic field. The interior product of
$\partial_{\bar{\theta}}$ and $\sigma$ is unit one, $\iota_{\partial_{\bar{\theta}}}\sigma=1$.
A natural question arises. Is $\sigma$ the covector of gyrophase
$d\theta$? Or, can $\sigma$ define the global gyrophase $\theta$?
To answer this question, we check the exterior derivative of $\sigma$ 

\begin{align}
d\sigma= & \frac{1}{2}\nabla\times\mathbf{R}\times d\mathbf{X}\wedge d\mathbf{X}-\left(\nabla R+\partial_{t}\mathbf{R}\right)\cdot d\mathbf{X}\wedge dt\:,\label{eq:dsigma}
\end{align}
where%
\begin{align}
\mathbf{R} & \equiv\nabla\mathbf{c}\cdot\mathbf{a}-\frac{1}{2}\left(\mathbf{b}\cdot\nabla\times\mathbf{b}\right)\mathbf{b}\,,\\
R & \equiv-\partial_{t}\mathbf{c}\cdot\mathbf{a}-\frac{1}{2}\mathbf{b}\cdot\nabla\times\mathbf{D}\,,
\end{align}
is a four-vector $\left(\mathbf{R},R\right)$ in the configuration
space. Since $d\sigma\neq0$, the dimensionless one-form $\sigma$
is not an exact form $\sigma\neq d\theta$, and we cannot define the
global gyrophase $\theta$ from $\sigma$. 

The traditional guiding center theory defines the gyrophase on a predefined
local orthogonal coordinate frame, which may not exist globally over
a non-trivial field topology. This issue was raised and discussed
by Sugiyama\citep{sugiyamaGuidingCenterPlasma2008,sugiyamaResponseCommentGuiding2009}
and Krommes\citep{krommesCommentGuidingCenter2009} in 2009. Soon
after, Burdy and Qin\citep{burbyGyrosymmetryGlobalConsiderations2012}
identified the obstruction to the global existence of gyrophase is
the vector field $\nabla\times\left(\nabla\mathbf{c}\cdot\mathbf{a}\right)$.
Bohosian \citep{boghosianCovariantLagrangianMethods1987} gave similar
results in his earlier work. However, the global gyro-phase is not
a necessary condition for gyro-symmetry. The gyro-symmetry depends
only on the homogeneity of the phase space within the range of Kruskal's
ring. The classical approach obscures the geometric meaning of the
gyro-symmetry. A coordinate-free geometric representation is a more
suitable alternative. The Eq.\prettyref{eq:dsigma} shows that the
existence of gyrophase depends on the integrability of $\sigma$,
or requires two-form $d\sigma=0$ to vanish everywhere. This is a
straightforward conclusion of the geometric method. 

Continuing the derivation of guiding center dynamics, the horizontal
Lagrangian one-form is gyro-independent
\begin{equation}
\bar{\eta}_{\parallel}=\left\langle \eta_{\parallel}\right\rangle =A+p_{\parallel}=\left(\mathbf{A}+\mathbf{V_{\parallel}}\right)\cdot d\mathbf{X}-\left(\phi+\frac{\mathbf{V_{\parallel}}^{2}}{2}-\mu\left|B\right|\right)dt\,,\label{eq:eta_parallel}
\end{equation}
which no additional calculations are required. Using Eq.\prettyref{eq:dsigma},
the Lagrangian two-form is rewritten as 

\begin{equation}
d\bar{\eta}=d\bar{\eta}_{\parallel}+d\bar{\eta}_{\perp}=\Omega^{\dagger}+d\mu\wedge\sigma\,,
\end{equation}
where %
\begin{align}
\Omega^{\dagger} & \equiv d\eta_{\parallel}+\mu d\sigma\nonumber \\
 & =\frac{1}{2}\mathbf{B}^{\dagger}\times d\mathbf{X}\wedge d\mathbf{X}+\mathbf{E}^{\dagger}\cdot d\mathbf{X}\wedge dt\nonumber \\
 & +\mathbf{b}\cdot d\mathbf{V}\wedge\mathbf{b}\cdot d\mathbf{X}-\left(\mathbf{V}-\mathbf{D}\right)\cdot d\mathbf{V}\wedge dt\,,\label{eq:Omega_dagger}
\end{align}
is the gyro-independent part of Lagrangian two-form, and%
\begin{align}
\mathbf{B}^{\dagger} & \equiv\mathbf{B}+\nabla\times\mathbf{V_{\parallel}}+\mu\nabla\times\mathbf{R}\,,\\
\mathbf{E}^{\dagger} & \equiv\mathbf{E}-\partial_{t}\mathbf{V_{\parallel}}-\nabla\left(\frac{\mathbf{V_{\parallel}}^{2}}{2}+\mu\left|B\right|\right)-\mu\left(\nabla R+\partial_{t}\mathbf{R}\right)\;,
\end{align}
are effective electromagnetic fields. Then, the Eq.\prettyref{eq:horizontal Hamiltonian equation}
is simplified as 

\begin{equation}
*\tau_{\parallel}^{\flat}=\frac{3!}{2!}\frac{d\mu\wedge\sigma\wedge\Omega^{\dagger}\wedge\Omega^{\dagger}}{\sigma\wedge\Omega^{\dagger}\wedge\Omega^{\dagger}\wedge\Omega^{\dagger}}=\frac{d\mu\wedge\sigma\wedge\Omega^{\dagger}\wedge\Omega^{\dagger}}{2\mathbf{B}^{\dagger}\cdot\mathbf{b}}\,.
\end{equation}
Substituting Eq. \prettyref{eq:Omega_dagger}, yields the Hamiltonian
vector field of guiding center %
\begin{align}
\tau_{\parallel t} & =\frac{\mathbf{B}^{\dagger}\cdot\mathbf{b}}{\left|B\right|}\,,\\
\tau_{\parallel\mathbf{X}} & =\frac{\mathbf{V}\cdot\mathbf{b}\mathbf{B}^{\dagger}+\mathbf{E}^{\dagger}\times\mathbf{b}}{\left|B\right|}-\mathbf{b}\times\nabla\mu\,,\label{eq:tau_X}\\
\tau_{\parallel\mathbf{V}} & =\left(\frac{\mathbf{B}^{\dagger}\cdot\mathbf{E}^{\dagger}}{\left|B\right|}+\mathbf{B}^{\dagger}\cdot\nabla\mu\right)\mathbf{b}\nonumber \\
 & +\left|\mathbf{V}_{\perp}\right|\left(\tau_{\parallel\mathbf{X}}\cdot\mathbf{R}-\tau_{\parallel t}R\right)\mathbf{a}\nonumber \\
 & -\frac{1}{2}\left|\mathbf{V}_{\perp}\right|\left(\tau_{\parallel\mathbf{X}}\cdot\nabla\ln\mu+\tau_{\parallel t}\partial_{t}\ln\mu\right)\mathbf{c}\,.\label{eq:tau_V}
\end{align}
where $\tau_{\mathbf{X}}$ is the drift motion of the guide center,
and $\tau_{\mathbf{V}}$ is the acceleration of Kruskal's ring. In
Eq. \prettyref{eq:tau_V}, the first term is the acceleration of the
guiding center in the direction of the magnetic field, the second
term represents the expansion of Kruskal's ring, and the third term
represents the rotational acceleration of Kruskal's ring. One can
simply verify that $\tau_{\parallel}$ is gyro-independent and orthogonal
to the gyro-motion $\mathcal{L}_{\partial_{\bar{\theta}}}\tau_{\parallel}=\iota_{\tau_{\parallel}}\eta_{\perp}=0$.
The magnetic moment $\mu$ is a constant for guiding center motion
$\iota_{\tau_{\parallel}}d\mu=0$.%

\section{Summary and discussion\label{sec:Summary}}

To summarize, we discuss the dynamics of a charged particle in a time-dependent,
slowly varying electromagnetic field, whose Lagrangian is a one-form
$\eta$ on the seven-dimensional contact manifold $P$. The orbit
of gyro-symmetry $\Phi_{\theta}$ is a closed ring in the phase space,
called Kruskal's ring $\mathcal{O}$. The guiding center is the center
of Kruskal's ring, the fixed point of gyro-symmetry $\Phi_{\theta}$.
By properly defining the rote-rate vector $\partial_{\bar{\theta}}$
(Eq.\prettyref{eq:rote-rate vector}) and the gyro-radius $\partial_{\bar{\rho}}$
(Eq.\prettyref{eq:rho}), we give a general expression for the gyro-averaging
(Eq. \prettyref{eq:averaged_eta_n} ). Further, the geometric decomposition
of the gyro-motion is given and verified (Eq. \prettyref{eq:eta_perp}
and \prettyref{eq:eta_parallel}). As a result, we obtain the coordinate-free
expression of the non-relativistic guiding-center dynamics in the
time-dependent slow-varying electromagnetic field. (Eq. \prettyref{eq:tau_X}
and \prettyref{eq:tau_V}).

We understand the gyro-symmetry as the similarity of the trajectory
of ringmates on the same Kruskal's ring, which depends only on the
local homogeneity of the electromagnetic field. The gyro-averaging
is integral along the Kruskal's ring. The guiding center is a particle
with \textquotedbl spin\textquotedbl , whose magnetic moment $\mu$
is equal to the action integral over the Kruskal's ring. The guiding
center dynamics is the dynamics of Kruskal's ring. The purpose of
guiding center theory is to decompose particle motion into vertical
gyro-motion and horizontal drift motion. In the classical guiding
center theory, the vertical part of Lagrangian is expressed as $\mu d\theta$.
We now recognize that $d\theta$ may not be globally defined on the
non-trivial magnetic field. Even if we ignore the global validity
and consider only the local domain, $d\theta$ is still an ambiguous
expression. Appling the local gyro-phase coordinate frame to the dimensionless
one-form $\vartheta$, yields $\vartheta=d\theta+d\mathbf{X}\cdot\nabla\mathbf{c}\cdot\mathbf{a}+dt\cdot\partial_{t}\mathbf{c}\cdot\mathbf{a}+\mathcal{O}\left(\varepsilon\right)$,
which means $d\theta$ only make scenes when the electromagnetic field
is homogeneous. The geometric decomposition method can avoid the confusion
caused by the local gyro-phase coordinates frame. The expansion (Eq.\prettyref{eq:expand_gamma})
and averaging method (Eq.\prettyref{eq:averaged_eta_n}) ensure that
each order of the expansion is gyro-independent.

In the subsequent work, the geometric methods established in this
paper will be applied to Lie perturbations, gyro-kinetics theory,
and others. Thus, the rich geometric of gyro-dynamics should be further
revealed.
\begin{acknowledgments}
The author sincerely thanks Prof. Chang-Xuan Yu, Prof. Wan-Dong Liu,
Prof. Jin-Lin Xie, and the unmentioned supervisors. The author is
grateful for their support during his early academic career. The author
also would like to thank Dr. Jian-Yuan Xiao and Dr. Pei-Feng Fan for
the fruitful discussions on geometry and numerical algorithm. 

This work was supported by the National MCF Energy R\&D Program under
Contract No.2018YFE0304102. 
\end{acknowledgments}

\bibliographystyle{unsrt}
\bibliography{main_submit}

\begin{thebibliography}{10}

\bibitem{caryHamiltonianTheoryGuidingcenter2009}
John~R. Cary and Alain~J. Brizard.
\newblock Hamiltonian theory of guiding-center motion.
\newblock {\em Reviews of Modern Physics}, 81(2):693--738, May 2009.

\bibitem{sugiyamaResponseCommentGuiding2009}
Linda E.~LE Sugiyama.
\newblock Response to {Comment} on "{Guiding} center plasma models in three
  dimensions".
\newblock {\em Physics of Plasmas}, 16(8):1--4, August 2009.

\bibitem{sugiyamaGuidingCenterPlasma2008}
Linda~E. Sugiyama.
\newblock Guiding center plasma models in three dimensions.
\newblock {\em Physics of Plasmas}, 15(9):1--13, September 2008.

\bibitem{krommesCommentGuidingCenter2009}
John~A. Krommes.
\newblock Comment on "{Guiding} center plasma models in three dimensions"
  [{Phys}. {Plasmas} 15, 092112 (2008)].
\newblock {\em Physics of Plasmas}, 16(8):084701, August 2009.

\bibitem{burbyGyrosymmetryGlobalConsiderations2012}
J.~W. Burby and H.~Qin.
\newblock Gyrosymmetry: {Global} considerations.
\newblock {\em Physics of Plasmas}, 19(5):052106, 2012.

\bibitem{burbyGuidingCenterDynamics2020}
J.~W. Burby.
\newblock Guiding center dynamics as motion on a formal slow manifold in loop
  space.
\newblock {\em Journal of Mathematical Physics}, 61(1):012703, January 2020.

\bibitem{burbyGeneralFormulasAdiabatic2020}
J.~W. Burby and J.~Squire.
\newblock General formulas for adiabatic invariants in nearly periodic
  {Hamiltonian} systems.
\newblock {\em Journal of Plasma Physics}, 86(6):835860601, December 2020.

\bibitem{burbyNormalStabilitySlow2021}
J.~W. Burby and E.~Hirvijoki.
\newblock Normal stability of slow manifolds in nearly periodic {Hamiltonian}
  systems.
\newblock {\em Journal of Mathematical Physics}, 62(9):093506, September 2021.

\bibitem{kruskalAsymptoticTheoryHamiltonian1962}
Martin Kruskal.
\newblock Asymptotic theory of {Hamiltonian} and other systems with all
  solutions nearly periodic.
\newblock {\em Journal of Mathematical Physics}, 3(4):806, July 1962.

\bibitem{qinShortIntroductionGeneral2005}
H~Qin.
\newblock A short introduction to general gyrokinetic theory.
\newblock In {\em Topics in kinetic theory}. PPPL, 2005.

\bibitem{omohundroGeometricPerturbationTheory1985}
S.~M. Omohundro.
\newblock {\em Geometric perturbation theory and plasma physics}.
\newblock Thesis/{Dissertation}, Lawrence Berkeley Lab., CA (USA), April 1985.

\bibitem{parraEquivalenceTwoIndependent2014}
F.~I. Parra, I.~Calvo, J.~W. Burby, J.~Squire, and H.~Qin.
\newblock Equivalence of two independent calculations of the higher order
  guiding center {Lagrangian}.
\newblock {\em Physics of Plasmas}, 21(10):104506, October 2014.

\bibitem{arnoldAppendixContactStructures1989}
V.~I. Arnold.
\newblock Appendix 4 {Contact} structures.
\newblock In {\em Mathematical {Methods} of {Classical} {Mechanics}}, volume~60
  of {\em Graduate {Texts} in {Mathematics}}. Springer New York, New York, NY,
  1989.

\bibitem{marsdenIntroductionMechanicsSymmetry1999}
Jerrold~E Marsden and Tudor~S Ratiu.
\newblock {\em Introduction to {Mechanics} and {Symmetry}}, volume~17 of {\em
  Texts in {Applied} {Mathematics}}.
\newblock Springer New York, New York, NY, 1999.

\bibitem{arnold44IntegralInvariang1989}
V.~I. Arnold.
\newblock 44 {The} integral invariang of {Poincare}-{Cartan}.
\newblock In {\em Mathematical {Methods} of {Classical} {Mechanics}}, volume~60
  of {\em Graduate {Texts} in {Mathematics}}. Springer New York, New York, NY,
  1989.

\bibitem{hallLieGroupsLie2015}
Brian~C. Hall.
\newblock {\em Lie {Groups}, {Lie} {Algebras}, and {Representations}: {An}
  {Elementary} {Introduction}}, volume 222 of {\em Graduate {Texts} in
  {Mathematics}}.
\newblock Springer International Publishing, Cham, 2015.

\bibitem{arnoldTheoryPerturbationsConditionally1989}
Vladimir~Igorevich Arnol'd.
\newblock Theory of perturbations of conditionally periodic motion.
\newblock In {\em Mathematical methods of classical mechanics}, page 405.
  Springer-Verlag, 1989.

\bibitem{feckoDifferentialGeometryLie2006}
Mari\{{\textbackslash}'a\}n Fecko.
\newblock {\em Differential {Geometry} and {Lie} {Groups} for {Physicists}}.
\newblock Cambridge University Press, Cambridge, 2006.

\bibitem{littlejohnHamiltonianPerturbationTheory1982}
Robert~G. Littlejohn.
\newblock Hamiltonian perturbation theory in noncanonical coordinates.
\newblock {\em Journal of Mathematical Physics}, 1982.

\bibitem{littlejohnGuidingCenterHamiltonian1979}
Robert~G. Littlejohn.
\newblock A guiding center {Hamiltonian}: {A} new approach.
\newblock {\em Journal of Mathematical Physics}, 1979.

\bibitem{caryNoncanonicalHamiltonianMechanics1983}
John~R Cary and Robert~G Littlejohn.
\newblock Noncanonical {Hamiltonian} mechanics and its application to magnetic
  field line flow.
\newblock {\em Annals of Physics}, 151(1):1--34, November 1983.

\bibitem{caryLieTransformPerturbation1981}
J~Cary.
\newblock Lie transform perturbation theory for {Hamiltonian} systems.
\newblock {\em Physics Reports}, 79(2):129--159, December 1981.

\bibitem{qinPullbackTransformationsGyrokinetic2004}
H~Qin and W~M Tang.
\newblock Pullback transformations in gyrokinetic theory.
\newblock {\em Physics of Plasmas}, 11(3), 2004.

\bibitem{jacksonClassicalElectrodynamics1999}
John~David Jackson.
\newblock {\em Classical electrodynamics}.
\newblock Wiley, New York, 3rd ed edition, 1999.

\bibitem{boghosianCovariantLagrangianMethods1987}
BM~M Boghosian.
\newblock {\em Covariant {Lagrangian} methods of relativistic plasma theory}.
\newblock PhD thesis, UNIVERSITY OF CALIFORNIA DAVIS, 1987.

\end{thebibliography}

\end{document}